\documentclass[12pt,a4paper]{article}
\usepackage{amssymb,amsfonts,amsmath}

\title{Reflections on Zeilinger-Brukner  information interpretation of quantum mechanics}

\author{Andrei Khrennikov\\
International Center for Mathematical Modelling \\
in Physics and Cognitive Sciences\\
Linnaeus University,
V\"axj\"o, SE-351 95, Sweden\\
Andrei.Khrennikov@lnu.se}

\date{}
 
\begin{document}

\maketitle

\begin{abstract} In this short review I present my personal reflections on Zeilinger-Brukner  information interpretation of quantum mechanics (QM). 
In general, this interpretation is very attractive for me. However, its rigid coupling to the notion of irreducible quantum randomness is a very complicated issue
which I plan to address in  more detail.   This note may be useful for general public interested in quantum foundations, especially because 
I try to analyze essentials of the information interpretation critically
(i.e., not just emphasizing its advantages as it is commonly done). This review is written in non-physicist friendly manner.  
Experts actively exploring this interpretation  may be interested in the paper as well, as in the comments of ``an external observer'' who have been  monitoring the
development of this approach to QM during the last 18 years. The last part of this review is devoted to the general methodology of science 
with references to views of de Finetti, Wigner, and Peres. 
\end{abstract}

\section{Introduction}

Tremendous development of quantum information, both theory and experiment, led to considerable growth of interest to foundational problems of QM. Moreover, nowadays quantum 
information approaches the stage of technological applications, with startups in quantum cryptography and quantum random generators. \footnote{ In particular, recently  three leading experimental 
groups \cite{Hensen}, \cite{Giustina}, \cite{Shalm}  claimed that they were able to perform the loophole free test of violation of Bell's inequality. This inequality plays of the role of borderline 
between classical and quantum physics. Experimental verification of its violation is the endpoint in the famous debate between Einstein and Bohr; it also  justifies  the mentioned  
quantum technological projects - quantum cryptography and  random generators (especially the latter).} 

This recent wave of foundational studies is not surprising, because the aforementioned success in theory and experiment is shadowed by the recognition that 
 the basic foundational problems of QM have not yet been solved, in spite of one hundred years of tremendous efforts of the brightest minds in physics and philosophy.
One of  these problems is the problem of elaboration of a consistent and commonly acceptable interpretation of QM, an interpretation of mathematical entities used 
in the quantum formalism.  Formal rules of mathematical manipulations are well defined and they function very well in all possible applications. However, there is still no consensus on 
 the physical meaning  of (at least) some of them. 

The modern situation in QM is characterized by the huge diversity of interpretations. 
Moreover, as was remarked in my book \cite{Beyond}  each of basic 
interpretations has a variety of individual flavors. For example, suppose that somebody 
claims to be an adherent of the Copenhagen interpretation of QM. When you ask for details, most probably
 you will hear a rather specific view on what  this interpretation
is about. I have the impression that there are so many  versions of the Copenhagen interpretation of QM as there are
people knowing about it. The same can be said about, e.g., the statistical or Bohmian interpretations of QM.  
This situation is unacceptable for any rigorous and well established theory. Therefore experts in foundations invest so much effort in the problem of interpretation.

Development of quantum information theory enlightened the fundamental role of information in QM and led to elaboration of a variety 
of {\it information interpretations} of QM. In this short review we would like to present the most widely recognized version - the information    
interpretation of A. Zeilinger and C. Brukner \cite{Z0}, \cite{BR1}- \cite{BR6}, \cite{Z1} (see also  Kofler and Zeilinger \cite{Kofler}
and Brukner, Zukowski, and Zeilinger \cite{BRZ}).   
 I  have been following development of this interpretation from its very beginning \footnote{Starting from conferences in Helsinki on quantum foundations in 1990th and conversations
with A. Zeilinger at a number of conferences at the beginning of this century to my recent discussions with A. Zeilinger and C. Brukner during the V\"axj\"o conferences and my visiting fellowships in Vienna,
} observing its evolution and success (quite natural in light of recent development of quantum information).    
My reflections on   the  information interpretation can be treated as ``external observer'' reflections. I hope that this presentation has some degree of objectivity (cf. with the above remark 
of the role of subjective factor in problem of interpretations of QM). 

The last part of this review is devoted to general scientific methodology with references to contributions of de Finetti, Wigner, and Peres.

\section{Zeilinger-Brukner  Information Interpretation}
\label{BZ77}

In the {\it information interpretation}\index{information interpretation}  of QM, information\index{information} is the most fundamental, basic entity. Every quantized system 
is associated with a definite discrete amount of information (see Zeilinger \cite{Z0}, also \cite{Z1}). This information content remains constant during evolution of a closed system. Here a quantum state
 is defined in the spirit of Schr\"odinger, see  \cite{Schr}:  
{\it the quantum state  is an expectation catalog (of probabilities for all possible outcomes).} We remark that Zeilinger elaborated the
 information interpretation of QM  \cite{Z0} by searching 
for a fundamental and heuristically  clear principle of QM, similar to {\it Einstein's principle of relativity.}\index{principle of relativity} 

A. Zeilinger who presented the basic principles of the information interpretation  in 1999 \cite{Z0} always  emphasized its close connection  
with the Copenhagen interpretation; in particular, he  often cited  
N. Bohr to emphasize connection with Bohr's ideas. The same line of presentation continues in joint publications  of Zeilinger and Brukner \cite{BR1}-\cite{BR6},
Kofler and Zeilinger   \cite{Kofler} and Brukner, Zukowski, and Zeilinger \cite{BRZ}. 
Indeed, the information interpretation of QM can be considered as a modern information-theoretic version of the orthodox
Copenhagen interpretation. It has some commonality with von Neumann's version of this interpretation.   In particular,  Zeilinger and Brukner explore heavily the concept of 
{\it irreducible quantum randomness} \index{irreducible quantum randomness} which  was invented by von 
Neumann \cite{VN}.\footnote{ Surprisingly they
 did not refer to von Neumann at all (of course, it might be that I missed some of their other papers with a corresponding reference to von Neumann as the inventor of 
irreducible quantum randomness). They use the term {\it ``objective randomness''}\index{objective randomness}, but the meaning of this term coincides with 
the meaning of von Neumann's irreducible randomness.} They also consider a quantum state as a state of an individual 
physical system.  

As all interpretations in the spirit of Copenhagen \cite{PL0}-\cite{PL_KHR},  the information interpretation is non-realistic (see also \cite{Jaeger}).
By this interpretation QM is not about features of quantum systems, but about  information about these systems gained 
with the aid of (classical) measurement devices. In \cite{PL_KHR} it was emphasized that one has to distinguish realism and reality. For example, 
Bohr used 
the non-realist interpretation of QM \cite{BR7}, \cite{BRF}, but he definitely did not deny reality of atoms, electrons (and later even photons). 
It is not clear whether the information interpretation needs reality - beyond outputs of measurement devices (e.g., here   photon is treated as  
a click of a detector). \footnote{In general,
Viennese do not like realism and even reality.  The following story represents perfectly  Viennese's viewpoint on reality. Once my friend Johan Summhammer (Atom Institute, Vienna) spent 
two weeks in V\"axj\"o. This town is surrounded by beautiful lakes (``sj\"o'' is a lake in Swedish). Once I met him looking at a lake, really excited. I asked him
about reason of the excitement, expecting a typical comment about beauty of Swedish nature. But, Johan answered that he is excited by this great picture created by clicks 
of detectors composing his brain.'  From long conversations with Johan I learned that there is no objective reality, and bio-systems developed an ability to select some patterns 
from noisy and unstructured environment and cognition was developed in this way. Thus this world is composed of clicks of detectors. May be not all Viennese  
share  this view of ``clicks-made reality'', but the above story defintely reflects the general Vennese attitute.  There is something in the spirit of the town...}

A. Zeilinger has put forward an idea which {\it connects the concept of information with the notion of elementary systems.}\index{elementary system} 
Here we follow the presentation from his joint paper  with J. Kofler \cite{Kofler}:

The description of the physical world is based on propositions. Any physical object can be described by a set of true propositions. 
Then Zeilinger pointed out that ``we have knowledge or information about an object only through observations.''\footnote{Here one of the important problems of this interpretations lies in assigning 
the meaning to ``we''.  ``Whose knowledge?'' as was \index{knowledge} asked by D. Mermin \cite{Mermin_KN}. We shall come back to this
 problem later.} Everybody would agree with this. But further line of reasoning is not unquestionable.  

{\small ``Any complex object which is represented by numerous propositions can be decomposed into constituent systems which need fewer propositions to be specified. 
The process of subdividing reaches its limit when the individual subsystems only represent a single proposition, and such a system is denoted as an elementary system.
 The truth value of a single proposition about an elementary system can be represented by one bit of information with ``true'' being identified with the bit value ``1'' and  ``false'' with  ``0''. ''}
It is then suggested to assume  

\medskip

 {\bf Principle of quantization of information:} {\it An elementary system carries one bit of information.}

\medskip

For me, the main questionable point of this reasoning is validity of such subdivision. Why it is always possible? Cannot it be the case that two 
bits of information belong to a system and cannot be separated? Anyway, the possibility of decomposition as described above looks more like another postulate. 
From the principle of  quantization of information, Zeilinger derives objectivity (irreducibility) of quantum randomness. (In contrast to von Neumann, he needs no mathematically 
complicated no-go theorem.) Further we again follow the paper of  Kofler and Zeilinger \cite{Kofler}:

{\small ``Disregarding the mass, charge, position and momentum of the electron, its spin is such an elementary system. 
If it is prepared ``up along $z$'', we have used up our single bit and a measurement along any other direction must necessarily contain an element of randomness. 
This randomness must be objective and irreducible. It cannot be reduced to some unknown hidden properties as then the system would carry more than a single bit of information. 
Since there are more possible experimental questions than the system can answer definitely, it has to ``guess''. Objective randomness is a consequence of the principle lack of information.''}

I appreciate the elegance of this derivation of irreducibility of quantum randomness. 
In contrast to von Neumann, Zeilinger has no need for conditions of coupling between mathematical formalisms of a possible subquantum model and  QM. His fundamental principle 
is formulated in heuristically clear terms, i.e., without any direct relation to, e.g., microworld. 
However, I stress again that Zeilinger's  basic principle of quantization of information is questionable. Other Copenhagenists (Bohr, Heisenberg, Pauli, von Neumann) proceeded without it
(as well as recently  Plotnitsky  \cite{PL_KHR} with his ``statistical Copenhagen interpretation'' ). 

Finally, we remark that in the derivation of irreducibility of quantum randomness there was encrypted the application 
of the {\it principle of complementarity}  in the form:   
"Since there are more possible experimental questions than the system can answer definitely...''. Thus Zeilinger's information interpretation is based on two fundamental principles:
\begin{enumerate}
\item  Principle of quantization of information.
\item Principle of complementarity.
\end{enumerate}
We note that if two observables (experimental questions) are not complementary, that is, can be asked simultaneously or in any order, it means, mathematically, 
that their operators commute. There is a theorem, liked by von Neumann \cite{VN},  that for commuting operators $A$ and $B$ 
a new observable $C$ can be constructed such that values of $A$ and $B$ can be unambigously extracted from the result of $C$ measurement, i.e. there are two functions 
$f=f(x)$ and $g=g(x), x \in \mathbf{R},$ such that $A=f(C)$ and $B=g(C).$
Hence,  {\it no complementarity means no inherent randomness.}

\medskip

As was pointed out by one of the reviewers, the principle of complementarity could be replaced by {\it an assumption that there are independent propositions --
 those which do not have joint true values.} This would already imply that there must
 exist complementarity,  as a consequence of the principle of quantization of information and the independence of propositions.

\medskip

Now it is really the time to turn to the questions: ``Whose knowledge? Whose information?'' I think that these are the ``hard questions'' of the information approach to QM.
Surprisingly here the positions of Zeilinger and Brukner do not coincide. 

In June 2014 at the conference in Vienna devoted to 50th anniversary of Bell's inequality, Zeilinger gave the 
talk presenting the  personal agent viewpoint on information encoded in a quantum state. The wave function is in the head and not in nature (von Neumann would definitely disagree, 
Bohr would probably agree, Wigner would applaud). And the main point is that it is in the {\it concrete head}, his head or my head. This viewpoint is close to the views of 
Fuchs,  Schack, Caves, Mermin, \cite{Caves1}-\cite{Fuchs6} (see also my recent critical review on QBism \cite{QBism_KHR} and even more critical paper of  
Marchildon \cite{Marchildon}).\footnote{The talk of Zeilinger generated very polarized reactions. For example, Aspect strongly commented that in the two slit experiment the photon ``knows'' from the 
very beginning that it is forbidden  to go to some regions on the registration screen, independently of what happens in Zeilinger's head. Then another provocative question was asked 
to Zeilinger:  Is it important to be a human being in order to have a wave function in the head? The reply was in the spirit that this is 
not important and dog's head is also a good machine 
to present a wave function! Of course, 
it might be that this was just a provocative joke response to this provocative question. 
If not, then additionally to the trouble with Schr\"odinger cat \index{Schr\"odinger cat} \index{Schr\"odinger cat} we got a new trouble - with a
Zeilinger's dog.\index{Zeilinger dog} I remark that all these strange creatures were 
born in Vienna. (But in general this discussion after Zeilinger's lecture reminds the koan about whether a six-trunk 
white elephant \index{white elephant} can have a Buddha \index{Buddha} nature.)}    This viewpoint on a quantum state as an information entity used in decision making\index{decision making} regarding experimentally observed probabilities 
 well matches interpretations of QM as the machinery for update of probabilities - QBism and the V\"axj\"o interpretation. I cannot confidently say how 
Zeilinger interprets probability: statistically or subjectively? During my lectures on foundations of probability at his seminar in May-June 2014 the subjective interpretation of probability 
was not mentioned at all. I used consistently the statistical interpretation and it seemed that all participants of the seminar were fine with this.  

At the same time recently Brukner published a paper on what I interpret as the universal (i.e., not private as in QBism)  
agent perspective on the information interpretation, we cite Brukner \cite{BR4}:  

{\small ``The quantum state is a representation of knowledge necessary for a hypothetical observer 
 respecting her experimental capabilities to compute probabilities of outcomes of all possible future experiments.'' }

Here an explicit reference to the observer's experimental capabilities is crucial, cf. 
with my analysis of QBism in \cite{QBism_KHR}, section : {\it Agents constrained by Born's rule.}

It has to be noted that in this paper the author emphasized the closeness to QBism. With this I strongly disagree. From the QBism perspective, the wave function 
is in the head of a concrete private agent, e.g., in Fuchs' head, not in the head of a hypothetical observer. Similarly, Zeilinger spoke (at least at the aforementioned 
occasion) about his private wave function (or even his dog's wave function, again the latter was may be just a joke, but later we shall discuss this point seriously), 
not the wave function used by a hypothetical observer. 

The ``knowledge'' here refers to Wigner's definition of the quantum state: 
{\small ``... the state vector is only a shorthand expression
of that part of our information concerning the past of the system which is relevant
for predicting (as far as possible) the future behavior thereof.''}

Unfortunately,  it seems that de Finetti and Wigner were not familiar with the works of each other.   And the reader can 
see \cite{Finetti}, \cite{Finetti1}  (see also \cite{QBism_KHR}) that de Finetti's viewpoint 
on probability - its subjective (personal knowledge) interpretation - is very close to Wigner's viewpoint on quantum state.   

As Brukner pointed out,  {\small ``Peres correctly notes that considering hypothetical observers is not a prerogative of quantum
theory [...]. They are also used in thermodynamics, when we say that a perpetual-motion machine
of the second kind cannot be built, or in the theory of special relativity, when we say that no signal
can be transferred faster than the speed of light.''}

This is a natural conclusion; it would be surprising if hypothetical observers and Wigner's interpretation \index{Wigner interpretation} of a state were applicable only in quantum physics.
We can refer to de Finetti   \cite{Finetti}, \cite{Finetti1} who starting with the subjective interpretation of probability developed a subjective experience methodology of science, see \cite{QBism_KHR} for detailed 
presentation. De Finetti, of course,   did not  reduce his methodology to  quantum physics at all. (However, in contrast to Brukner,  de Finetti advertised the private and not universal agent perspective;
so Fuchs et al. (and, it seems, Zeilinger as well) are closer to de Finetti than Brukner).

In the light of this statement of Peres (completed by views of de Finetti, Wigner, Zeilinger, Brukner) the following question naturally arise: 
What are distinguishing features of  ``information concerning the past of the system which is relevant
for predicting'' related to quantum systems? Why does information about them is represented mathematically in such a special way, by vectors of complex 
Hilbert space?   De Finetti's framework   \cite{Finetti}, \cite{Finetti1}  (see also \cite{QBism_KHR}) covered homogeneously all areas of science (natural and humanities);  in the same way Wigner's viewpoint of a state as a collection of information 
is applicable to both classical and quantum states of both biological and physical systems. How can  the general  framework of de Finetti
be reduced to the special quantum representation?

One can refer to  the principle of quantization of information as the key principle restricting de Finetti's general viewpoint on scientific method; so to say,  

\medskip

{\it ``classical systems'' are those where  we are not able to approach the level of a single proposition description, the single bit level. }

\medskip

This viewpoint  presumes the possibility to derive the quantum (complex Hilbert space)  representation from this principle. And we point that 
there are already several complete reconstructions of QM from information-theoretical principles \cite{R1}--\cite{R1az}, \cite{BR3}; we point especially 
to the  work of Dakic and  Brukner \cite{R1b} as their axiom 1 is based on Zeilinger's principle of quantization.
\footnote{This problem also was understood well by QBists who tried to reconstruct the quantum formalism from their fundamental principle: QM is a special machinery 
for probability (information) update based on the special nonclassical version of the formula of total probability \cite{QBism_KHR}. But QBists succeeded only partially in approaching this great aim.
The same problem was actively studied in development the V\"axj\"o interpretation of QM which is based on the same principle as QBism, but explores another version of this formula,
see \cite{QBism_KHR}, \cite{INT_KHR}-\cite{KHR_CONT} for details. Nor this interpretation managed to solve the problem of 
quantum reconstruction.}    

I think that to explain peculiarity of the quantum representation of information one has to discuss the logical structure of information processing by humans. Besides classical Boolean logic,
there exist various nonclassical logics, nonclassical rules for information processing. In spite of his revolutionary treatment of the concept of probability and the scientific method in general,
de Finetti was still rigidly devoted to Boolean logic (and hence to the use of measure-theoretic probability concept). In spite of  wide applications of Boolean logic, e.g., in artificial intelligence and computer science,
we do not forget that it is just one special model of information processing which was created by a concrete person. Meanwhile, the brain may use more complex logical systems. In particular, it may 
use {\it quantum logic.} Thus the quantum representation of information  is a mathematical signature of the use of quantum logic in reasoning. Of course, from 
this viewpoint the context of quantum mechanical reasoning is only a special context in which the brain uses nonclassical logic. Such 
nonclassical reasoning may be profitable for the brain in other 
situations, see \cite{UB_KHR} on applications of the quantum formalism in cognitive science, psychology, economics.

Now we come back to Zeilinger's dog issue. Applying the quantum formalism outside of physics we discover \cite{UB_KHR}, \cite{BOOK}, \cite{BIO}
 that nonclassical processing of information  is a feature not only of humans, 
but of all bio-systems, from cells and proteins to, e.g.,  dogs. From this viewpoint, Zeilinger's dog also processes information (in some situations) by using nonclassical logic of reasoning and 
hence (roughly speaking) has wave functions in its head. Of course, the above attempt to couple the information interpretation to general theory of  reasoning and decision making 
is my own speculation; it has nothing to do with the views of 
adherents of this interpretation (see \cite{BIO} for further discussions).
       
I would like to thank A. Zeilinger, C. Brukner, K. Svozil for critical discussions on the information interpretation and hospitality during my visits to Vienna.
This study was supported by the grant ``Mathematical Modeling of Complex Hierarchic Systems'' of Linnaeus University and the grant Quantum Bio-Informatics of Tokyo 
University of Science.

\end{document}